\documentclass[pre,twocolumn,aps,superscriptaddress,showpacs]{revtex4}
\usepackage{epsfig}
\usepackage{bm}

\newcommand{\vp}{{\bf  v}}
\newcommand{\vf}{{\bf  u}}

\newcommand{\br}{{\bf r}}
\newcommand{\BE}{\begin{equation}}
\newcommand{\EE}{\end{equation}}
\newcommand{\beqn}{\begin{eqnarray}}
\newcommand{\eeqn}{\end{eqnarray}}
\newcommand{\lapla}{{\bf \nabla}}

\begin{document}

\title{Continuum description of finite-size particles advected by external flows.
The effect of collisions.}

\author{Crist\'obal L\'opez}
\affiliation{
Instituto Mediterr\'aneo de Estudios Avanzados
IMEDEA (CSIC-UIB) and Departament de Fisica, 
Universidad de las Islas Baleares,
E-07071 Palma de Mallorca, Spain.}
\author{Andrea Puglisi}
\affiliation{
INFM Center for Statistical Mechanics and Complexity and
Dipartimento di Fisica, Universit\`a di Roma `La Sapienza',  Piazzale A. Moro 2, I-00185, Rome,
Italy.}
\date{\today}

\begin{abstract}
The equation of the density field  of an assembly
of macroscopic particles advected by a hydrodynamic flow
is derived from the microscopic description of the system. This
 equation allows to recognize the role and the relative
importance of the different microscopic processes implicit in the
model: the driving of the external flow, the inertia of the particles,
and the collisions among them.
 The validity of the density description is
confirmed by comparisons of numerical studies of the
continuum equation
with
Direct Simulation Monte Carlo (DSMC)
simulations of hard disks advected by a chaotic
flow. We show
that the collisions have two competing roles: a dispersing-like
effect and a clustering effect (even for elastic collisions). An
unexpected feature is also observed in the system: the presence of collisions can reverse
the effect of inertia, so that grains with lower inertia are more
clusterized.
\end{abstract}
\maketitle

\section{Introduction}

The transport of inertial particles in hydrodynamic flows has recently
attracted the attention of many researchers~\cite{general}. The great
number of applications of this topic, including for e.g.  ecological
problems (dynamics of plankton populations), geophysical processes
(cloud formation, transport of pollutants in the atmosphere or the
oceans), or chemical engineering (chemical reactions in turbulent
or chaotic
flows), constitutes the main reason for this interest.

In most of these works, despite of considering finite-size particles,
the effect of the colliding processes among particles was completely
disregarded.  However, in some applications, as exemplified by rain
initiation in clouds~\cite{falkovichnature}, the role of collisions
seems to be crucial.  Thus, in order to take them into account, in a
recent paper~\cite{pre} we have introduced a simple model of a
granular material, $N$ particles subject to mutual collisions (elastic
or inelastic), advected by a two-dimensional chaotic flow (gravity is
not considered).  The model equivalently describes inertial particles
colliding among them and immersed in a chaotic flow, with the density
of the particles higher than that of the fluid.  A novel result found
in~\cite{pre} is that collisions may strongly modify the scenario of
the so-called {\it preferential concentration}~\cite{general} by which
particles in turbulent flows, in the absence of collisions, tend to
aggregate in specific spatial areas.

In this Paper we derive from first principles the equation for the
macroscopic density of grains (we indistinctly speak of grains or
colliding particles) for the model of~\cite{pre} (but with a flow 
not neccesarily chaotic).  Its properties are
discussed and, in particular, we show how it can explain two different
effects of collisions: dispersion and clustering.  In adition, the
macroscopic equation helps to understand another surprising feature of
the model, observed in the Direct Simulation Monte Carlo (DSMC)
simulations, that we call {\it reversed clustering}: in the presence
of collisions lower values of inertia can induce more aggregation of
the particles.

The article  is organized as follows: in the next section we present our 
model and derive an
explicit expression for the velocity of the grains. In Sec.~\ref{sec:density}
we obtain the evolution equation for the density field of particles. Next,
in Sec.~\ref{sec:discussion} we present some of the most relevant features
of this density equation; in particular, we discuss the relative importance
of its different terms. In Sec.~\ref{sec:numerics} we numerically check 
our theoretical results, and finally, in Sec.~\ref{sec:summary} we 
summarize the Paper.

\section{Equations of motion and stochastic treatment of collisions}

The equations of motion  of $N$ granular
particles driven by an external velocity field $\vf (\br,t)$ are given
by
\begin{eqnarray}
&\frac{d\vp_i (t)}{dt}=-\frac{1}{\hat{\tau_p}}\left( \vp_i (t)-\vf (\br_i(t),t)\right)&
\nonumber \\
&+\gamma \sum_{j=1}^N\sum_{k}\left[\left(\vp_i(t)-\vp_j(t)\right)\cdot \hat {\bf n}
_{ij}(t)\right]
\hat{\bf  n}_{ij}(t)\delta (t-t_{ij}^k),&
\label{modelo1} \\
&\frac{d\br_i (t)}{dt}=\vp_i (t), &
\label{modelo2}
\end{eqnarray}
where $i=1,...,N$, $\vp_i$ is the velocity of particle $i$, $\br_i$
its position, $\hat{\tau_p}$ is the Stokes time, $\gamma=(1+r)/2$,
$r\in[0,1]$ is the restitution coefficient ($r \leq 1$, with $r=1$ for
elastic collisions), $\hat {\bf n}_{ij}(t)$ is the unitary vector
joining the centers of particles $i$ and $j$ at time $t$ (in the
following we will also use the notation $\hat {\bf n} (\br,\br')$ to
denote a unitary vector joining vectors $\br$ and $\br'$), $\delta$ is
the Dirac delta, and with $t_{ij}^k$ we denote the times at which
particles $i$ and $j$ make their $k$-th collision. The first term in
eq.~(\ref{modelo1}) indicates the inertia of the particles, and the
second one shows how the velocity of a particle is modified because of
the collisions with the rest.
Without collisions eqs.~(\ref{modelo1}-\ref{modelo2}) are the equations of motion
of $N$ spherical particles where the Bernouilli term, the added-mass term,
the Basset-Boussinesq history integral term and the Faxen corrections are
neglected. Thus, 
 the advection of the
particles is modelized taking only the Stokes drag as the relevant
term in the Maxey and Riley equations~\cite{maxey}. This is a consistent 
approximation when the particles of the granular system  are much
heavier than those of the fluid~\cite{maxeyfluid}, which is the case that we are considering.
It is also fundamental
to mention that the relevant time scales of the problem were
identified in~\cite{pre} to be the typical time of the flow, $T$, the Stokes time,
$\tau_p$, the mean collision time, $\tau_c$, and, in the case of a chaotic
flow, the
time $\tau_f$, given by the inverse of the Lyapunov exponent of the
 flow.

In order to obtain a close expression for 
 $\vp_i (t)$ we first non-dimensionalize time and
velocities taking as unit scales the typical time $T$ and typical length
$L$ of the flow, i.e. $t \to t/T$, $u \to uT/L$ and $v \to vT/L$. Then
$\vp_i (t)$ will be obtained as a formal expansion in the non-dimensional
parameter $\tau_p=\frac{\hat{\tau_p}}{T}$. 
From eq.(\ref{modelo1}) on has straightforwardly
\begin{eqnarray}
&\vp_i (t)=\vf (\br_i (t),t)- \tau_p \frac{d\vp_i}{dt}&\nonumber \\
&-\gamma \tau_p
\sum_i \sum_k \left[ \left( \vp_i -\vp_j  \right)\cdot
\hat {\bf n}_{ij} (t) \right] \hat {\bf n}_{ij}(t) \delta (t-t_{ij}^k),& \nonumber \\
\label{trivial}
\end{eqnarray} 
then this expression of $\vp_i$ is itself substituted in the r.h.s of (\ref{trivial})
and one obtains to order   $\tau_p^2$ 

\begin{eqnarray}
\vp_i (t)&=&\vf (\br_i (t),t)- \tau_p \frac{d\vf}{dt}\nonumber \\
&-&\gamma \tau_p
\sum_i \sum_k \left[ \left( \vf (\br_i)-\vf (\br_j) \right)\cdot
\hat {\bf n}_{ij} (t) \right] \hat {\bf n}_{ij}(t) \delta (t-t_{ij}^k) \nonumber
\\
&+&\tau_p^2 \hat{\bm \eta}_i(t)+\Theta (\tau_p^3).
\label{trivial2}
\end{eqnarray}

If one writes down explicitly $\hat{\bm \eta}_i(t)$
\begin{eqnarray}
\hat{\bm \eta}_i(t)&=&\frac{d^2\vf (\br_i (t),t)}{dt^2} \nonumber \\
&+&\gamma \sum_j \sum_k
\huge(\frac{d}{dt}\left[\left(\vf (\br_i)-\vf (\br_j)\right) \delta (t-t_{ij}^k)\right]
\nonumber \\
&+&\delta (t-t_{ij}^k)\frac{d}{dt} \left(\vf (\br_i)-\vf (\br_j)\right)\huge)
\nonumber \\
&+&\gamma^2 \sum_j\sum_k \delta (t-t_{ij}^k)\huge(\sum_l \sum_{m}
(\vp_i-\vp_l)\delta (t-t_{il}^m)\nonumber \\
&-&\sum_n \sum_{p}(\vp_j-\vp_n)\delta(t-t_{jn}^p)
\huge),
\label{gorda}
\end{eqnarray}
one can see that it is 
a very complicated expression taking into account collisions among different particles.
Neglecting the first term in eq.~(\ref{gorda})
one can assume that it is some kind of effective stochastic term acting
at the colliding times (this hypothesis will be checked in Sec.~\ref{sec:numerics}),
 i.e., it denotes a random kick on a  particle every 
time it collides, 
\begin{equation}
\hat {\bm  \eta}_i (t)=\sum_{j=1}^N \sum_k
\delta(t-t_{ij}^k){\bm \zeta}_j (t),
\end{equation}
where ${\bm \zeta}_j (t)$ is a Gaussian white noise with  zero mean
and $\langle  \zeta^l_j (t)  
\zeta^m_k (t') \rangle=2D'\delta (t-t')\delta_{jk} \delta_{lm}$
(the superindex denotes vector component).

It is now easy to obtain that
$\langle \hat{\bm  \eta}_i (t) \rangle =0$,  and the correlations 
\begin{eqnarray}
&\langle  \hat{\eta}^k_i (t) \hat{\eta}^l_j (t') \rangle =
2D''\delta_{kl}\delta(t-t')\delta_{ij} 
\sum_j\sum_{m} \delta(t-t_{ij}^{m}) &\nonumber \\
&=2D \delta_{ij}\delta_{kl}\delta(t-t')n(\br_i(t),t),&
\label{correlaciones}
\end{eqnarray}
where $n(\br_i(t),t)$ is such that $n(\br,t)d\br dt$ is the average
 number of collisions that a particle
positioned in $\br$ undergoes at time $t$.
 $D'$ and
$D''$ are constants absorbed at the end in the definition of $D$, which is
another constant of order $\sigma^4/T^4$ ($\sigma$ is the diameter of the particles).
This last statement comes from comparing the full expression at order
$\tau_p^2$ in (\ref{gorda}) with (\ref{correlaciones}).

Finally, denoting $\hat{\bm \eta}_i(t)=\sqrt{D \ n(\br_i(t),t)} 
{\bm  \eta}_i (t)$ we arrive
to 
\begin{eqnarray}
\vp_i (t)&=&\vf (\br_i (t),t)- \tau_p \frac{d\vf}{dt}\nonumber \\
&-&\gamma \tau_p
\sum_i \sum_k \left[ \left( \vf (\br_i)-\vf (\br_j) \right)\cdot
\hat {\bf n}_{ij} (t) \right] \hat {\bf n}_{ij}(t) \delta (t-t_{ij}^k) \nonumber
\\
&+&\tau_p^2  \sqrt{D\ n(\br_i(t),t)} {\bm \eta}_i (t) +\Theta (\tau_p^3),
\label{velocidad}
\end{eqnarray}
where ${\bm \eta}$ is a Gaussian white noise with zero mean and
correlations $\langle \eta^k_i(t)
\eta^l_j(t')\rangle=2\delta_{kl}\delta_{ij}\delta(t-t')$. 

At this point we make the hypothesis that eq.~(\ref{velocidad}) is valid for
any $\tau_p$; in some sense, the 
higher order terms, $\Theta (\tau_p^3)$, just renormalize
the diffusion coefficient. This is supported by the fact that for large $\tau_p$ 
the noise term in eq.~(\ref{velocidad}) dominates, which is consistent with the fact
that in eq.~(\ref{modelo1}) for $\hat{\tau_p}$  large the dynamics of the particles
is mainly driven  by collisions. Thus, this hypotheses finally states that the 
net effect
of the colliding processes is properly modelized by the noise term appearing in
(\ref{velocidad}). This will also be checked in Section  V.

\section{Evolution equation of the density field}
\label{sec:density}

Our aim is to obtain the evolution equation of the density field of particles.
For this we closely follow~\cite{dean,umbertotarazona} and define
${\bar \rho} (\br,t)=\sum_{i=1}^N \rho_i(\br,t)\equiv \sum_{i} \delta (\br_i(t) -\br)$.
Using Ito Calculus~\cite{dean}
 
\begin{eqnarray}
&\frac{\partial \rho_i(\br,t)}{\partial t}= -\lapla \cdot (\rho_i \vf(\br,t))
+\tau_p \lapla \cdot (\rho_i \frac{d\vf}{dt}) &  \nonumber \\
&+\tau_p^4\frac{D}{2}\lapla^2 (n(\br,t)
\rho_i)
-\gamma \tau_p \lapla \cdot \huge[ \rho_i  \sum_{j=1}^N\sum_k & \nonumber \\
&\left[\left(\vf(\br,t)-\vf(\br_j(t),t)\right)\cdot \hat{\bf n}(\br,\br_j)\right]
\hat{\bf n}(\br,\br_j) \
\delta(t-t_{ij}^k)\huge]& \nonumber \\
&- \tau_p^2\sqrt{D}\lapla \cdot\left[\rho_i\sqrt{n(\br,t)}\ {\bm \eta}_i (t)\right].&
\label{deriv1}
\eeqn

We assume now that the local free time between collisions $\tau_c (\br,t)$ 
is everywhere smaller than any other characteristic time of
the system, and  that the mean free path $\lambda_c (\br,t)$  is also
smaller than any other characteristic spatial scale.
With these hypothesis we will proceed by integrating every term 
in (\ref{deriv1})
over small space-time cells $\Delta V \Delta t$ (with 
this integral normalised by this spatiotemporal volume)
 such that
$\tau_c(\br,t) <\Delta t< \min\{\tau_p, 1, \tau_f\}$, and
$\lambda_c^2(\br,t) <\Delta V<\lambda^2$,
with $\lambda$ the minimum typical space scale of the system. Assuming
that the fields ${\bar \rho}$ and $\vf$ are constant over the above
space-time cells, the integration over $\Delta V \Delta t$ (normalised
with this volume) and the
summation over index $i$ of the l.h.s in (\ref{deriv1}), and of the
first two terms in the r.h.s can be easily calculated.

More complicated are the other terms. Let us study them in detail beginning
with the diffusion term (the third in the r.h.s of (\ref{deriv1})):
\beqn
&\sum_i \frac{1}{\Delta V \Delta t} 
\int_t^{t+\Delta t}dt'\int_{\br}^{\br+\Delta V}d\br '
\frac{\tau_p^4 \ D}{2} \lapla^2(\rho_i(\br', t') \ n(\br' ,t'))  \approx & 
\nonumber \\
&\frac{\tau_p^4 \ D}{2\ \Delta V \Delta t} \sum_i \lapla^2 (\rho_i(\br,t)
\int dt' d\br ' n(\br ', t'))& \nonumber \\
&=\frac{\tau_p^4 \ D}{2\tau_c}\lapla^2 {\bar \rho}(\br,t)^2.& 
\label{difusion}
\eeqn
The first equality arises from the assumption that $\rho_i$
is constant in $\Delta V \Delta t$, and in the last one 
we have used that $\int dt' d\br ' n(\br ',
t')\approx {\bar \rho}(\br,t) \frac{\Delta V \Delta t}{\tau_c}$, which
follows from the above mentioned definition of the quantity 
$n(\br,t)d\br dt$.
 Note also
that we have assumed that $\tau_c$ is constant but all our results can
be extended to a $\tau_c$ dependent on the coordinates and time.

Next we proceed with the collision term, the fourth
one in the r.h.s of Eq~(\ref{deriv1}). Using the notation
${\bf s}(\br ', \br '', t') \equiv \left[ (\vf(\br ',t')-
\vf(\br '',t'))\cdot \hat {\bf n}(\br ' , \br '')\right] \hat {\bf n}(\br ' , \br '')$,
and, again,  that $\rho_i$ is approximately constant,
the sum over $i$ and integration over $\Delta V \Delta t$ can be
written
\beqn
&-\sum_i \frac{\gamma \tau_p}{\Delta V \Delta t} \int_{\Delta t}dt'
\int_{\Delta V}d\br '
 \lapla_{\br '} \cdot  \huge[\rho_i (\br ', t') & \nonumber \\
& \sum_{j,k} 
{\bf s}(\br ', \br_j(t'), t') \delta(t'-t_{ij}^k)\huge] \approx 
-\sum_i \frac{\gamma \tau_p}{\Delta V \Delta t} \lapla_{\br} \cdot \huge[& \nonumber \\
&\rho_i (\br, t) 
\int_{\Delta t}dt'\int_{\Delta V}d\br '
\sum_{j,k}
{\bf s}(\br ', \br_j(t'), t') \delta(t'-t_{ij}^k)\huge].
\end{eqnarray}
Then, 
evaluating the time integral we obtain
\begin{equation}
-\sum_i \frac{\gamma \tau_p}{\Delta V \Delta t}
\lapla_{\br} \cdot [\rho_i (\br, t) 
\int_{\Delta V}d\br '
\sum_{<j>} \sum_{<k>} 
{\bf s}(\br ', \br_j(t_{ij}^k), t_{ij}^k) ].
\end{equation}
Here the $\sum_{<k>}$ indicates a restriction in the sum to  the colliding
times $t_{ij}^k$ in the time interval $[t, t+\Delta t]$, and the
notation $\sum_{<j>}$ restricts the sum to the particles whose position is given by
$\br_j(t_{ij}^k)=\br ' +\sigma {\bf \hat n}_j$, where,
as already indicated,  $\sigma$ is the diameter of the
particles and ${\bf \hat n}_j$ is a unitary vector.
This calculation deserves some clarifications:  the restricted sum to
the colliding times appears because of the delta function, and the sum
restricted to the particles with the position as mentioned 
comes from the fact that, right at the collision event, two particles 
are separated
by a distance $\sigma {\bf \hat n}$, where ${\bf \hat n}$ is
an arbitrary unit vector.

Then we approximate $\sum_{<k>}{\bf s}(\br ', \br_j(t_{ij}^k),
t_{ij}^k) \approx
\frac{\Delta t}{\tau_c} {\bf s}(\br ', \br_j(t), t)$ because
$\frac{\Delta t}{\tau_c}$ is approximately the number of colliding
events in the time interval $\Delta t$,
and the  sum over $<j>$, with the help of the properties
of the Dirac delta function,  can be written as
\beqn
&\sum_{<j>} {\bf s}(\br ', \br_j(t), t) =& \nonumber \\
&\sum_j \int_{|{\bf \bar n}|=1} d{\bf \bar n} \
{\bf s}(\br ', \br '+\sigma  {\bf \bar n}, t)
\delta (\br_j (t)-(\br ' +\sigma {\bf \bar n}))\sigma, \nonumber \\
\label{deltadirac}
\end{eqnarray}
so that now we do not have to manage with a restricted sum. Moreover, the 
sum of the delta's is  the definition of ${\bar \rho}$, and we have that
(\ref{deltadirac}) is just
\begin{eqnarray}
&\sigma \ \int_{|{\bf \bar n}|=1} d{\bf \bar n}  \ {\bf s}(\br ', \br '+\sigma  {\bf \bar n}, t)
{\bar \rho} (\br '+\sigma {\bf \bar n}) & \nonumber \\
&= \sigma^2 \ \int_{|\bar n|=1} d{\bf \bar n}
[({\bf \bar n} \cdot \lapla_{\br'} \vf(\br',t)\cdot {\bf \bar n}]{\bf \bar n} {\bar \rho} (\br',t) 
+\Theta (\sigma^3);& \nonumber \\
\label{taylor}
\eeqn
the  equality comes from considering a Taylor expansion in powers of 
 $\sigma$.
Therefore, the collision term takes the form (assuming again constancy of the density
and velocity fields in $\Delta V$)
\BE
-\frac{\gamma \sigma^2 \tau_p}{\tau_c} \lapla_{\br}\cdot \left[
{\bar \rho} (\br,t)^2 \int_{|{\bf \bar n}|=1} d{\bf \bar n} 
[({\bf \bar n} \cdot \lapla_{\br} \vf(\br,t)\cdot {\bf \bar n}]{\bf \bar n} \right].
\label{colterm}
\EE

Finally, taking the average over the noise as in~\cite{umbertotarazona}, the 
last 
term in (\ref{deriv1}) disappears, 
 and noting 
$ \rho(\br,t) \equiv \langle {\bar \rho}(\br,t)\rangle$ we obtain the 
evolution equation for this density field~\cite{dosdim}
\beqn
&\frac{\partial  \rho(\br,t)}{\partial t}= -\lapla \cdot (\rho \vf(\br,t))
+\tau_p \lapla \cdot ( \rho \frac{d\vf}{dt}) &  \nonumber \\
&+\frac{\tau_p^4\ D}{2\tau_c}\lapla^2 (\rho^2) &\nonumber \\
&-\frac{\gamma  \sigma^2 \tau_p}{\tau_c} \lapla_{\br}\cdot \huge[
\bar \rho^2 \int_{|{\bf \bar n}|=1} d{\bf \bar n} 
[({\bf \bar n} \cdot \lapla_{\br} \vf(\br,t)\cdot {\bf \bar n}]{\bf \bar n} \huge]&.
\label{final}
\eeqn
Here we have used the  approximation 
$\langle \bar \rho^2\rangle \approx 
\langle  \bar \rho\rangle^2$, since we expect that
the densities in a point of space and time are likely to be
uncorrelated for different realizations of the noise. 

It is very important to clarify the meaning of the noise average
performed to obtain the $\rho$ field. This has been done following the
arguments of \cite{umbertotarazona}.  In our case, taken the average
over the noise esentially means smoothing out the higher order
corrections to the velocity of the particles coming from complicated
collision processes that take place in the time interval $\Delta t$.

\section{Relative importance of the terms in the density equation, and reversed clustering}
\label{sec:discussion}

Some relevant features of Eq. (\ref{final}) come inmediately to light. 
First, the inelastic character (the value of $\gamma$)
of the collisions is almost irrelevant 
at this level of description. The presence of the external 
driving flow  turns negligible the difference between elastic and inelastic
colliding processes. Also, the collision-induced diffusivity is, as expected,
dependent of the density of particles.
 This is clearly recognized by rewriting it as
$\frac{\tau_p^4\ D}{2\tau_c}\lapla^2 (\rho^2)=\frac{\tau_p^4\ D}{\tau_c}\lapla
(\rho  \lapla \rho) \equiv \lapla (\hat D(\rho)\nabla \rho)$,
where we have defined $\hat D(\rho)=\frac{\tau_p^4\ D}{2\tau_c} \rho$.

Most of the results of the discrete model (\ref{modelo1}-\ref{modelo2}) that were
presented in \cite{pre} can be understood in the framework of
Eq.~(\ref{final}). Let us put a label on each term of the r.h.s
of~(\ref{final}): the first is the pure advective term, which we label
with the letter $a$, the second is the inertial term, $b$ (of order
$\tau_p$), which is responsible for preferential concentration in the
absence of collisions~\cite{general}, the third is a
collision-induced diffusive term, $c$, of order $\tau_p^4 D/\tau_c$,
and the last is a collision-induced term, $d$, of order $\tau_p
\sigma^2/\tau_c$, that could also induce clustering.  

In the absence of collisions ($\tau_c \to
\infty$) the last two terms vanish and there is only competition
between the  advection, $a$, and the clusterizing
inertial term $b$: for e.g. when $\tau_p > \sim  1$ it appears that
inertia dominates and the system is strongly clusterized. On the other
side, in the limit $\tau_p \to 0$ ($b$ vanishes), it is
the $d$ term that can induce clustering (when $\tau_c \ll
\tau_p$); however the term $c$ may eventually become
stronger than $d$ giving rise to a homogenous distribution of
particles. The unexpected feature emerging from this analysis is that
the diffusive homogenizing term $c$ dominates at high values of 
$\tau_p$, thus {\it reversing} the effect of inertia that, in
collision-less systems, enhances concentration. In the presence of
collisions, therefore, we have a {\it reversed clustering} phenomenon:
for small values of $\tau_p$, i.e. inertia, 
the density is more clusterized than at larger
values.

\begin{figure}[htb]
\begin{center}
\includegraphics[clip=true,width=\columnwidth, keepaspectratio]{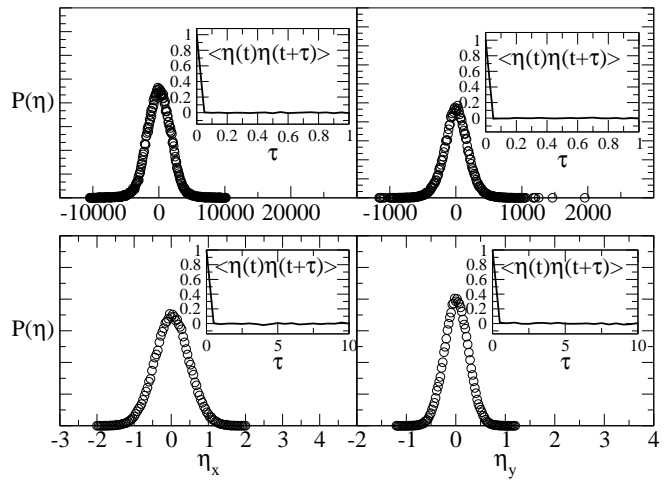}
\end{center}
\caption{
Distributions of the residual term (left: $x$ component, right: $y$
component) of eq.~(\ref{velocidad}) which is assumed to be a white
Gaussian noise. The form of the distribution 
 and the self-correlation (displayed in the insets)
confirm this assumption. The flow used in this simulation is a
sinusoidal shear, i.e.: $u_x(\mathbf{r}_i(t),t)=U\cos(2\pi y_i(t)/L)$,
$u_y=0$, where $\mathbf{r}_i(t)=(x_i(t),y_i(t))$.  For the upper plot
(large $\tau_p$) the time step used in the DSMC is $\Delta t=0.0001$,
size of the DSMC cells is $0.6 \times 0.6$, mean free time is $0.01$,
mean free path $5$, length of the time period used to calculate $\eta$
is $\overline{\Delta t}=0.005$, $\hat{\tau_p}=1$, $r=1$, $N=L^2=1000$,
$\sigma=0.3$, $U=10000$ so that $\tau_p=\frac{U\hat{\tau_p}}{L}\sim
300$.  For the bottom plot (small $\tau_p$) $\Delta t=0.01$, DSMC cells
are $2 \times 2$, mean free time is $0.6$, mean free path $18$, $\overline{\Delta
t}=0.5$, $\hat{\tau_p}=1$, $r=1$, $N=L^2=10000$, $\sigma=0.6$, $U=50$ so that
$\tau_p=0.5$.  }
\label{fig:stochastic}
\end{figure} 

\section{Numerical simulations}
\label{sec:numerics}

We have checked numerically our system in two spatial dimensions by
means of DSMC simulations of the hard disks model (similar to those
performed in
\cite{pre}) and by a numerical study (a Lax integration scheme) of
eq.(\ref{final}). The DSMC is a well established algorithm~\cite{dsmc}
that allows to simulate gaseous systems in a rapid and efficient way,
with the assumption of negligible correlations at short range.  We
have used a variant of DMSC that takes into account the Enskog factor
due to high density corrections, in order to obtain more accurate
simulations of the clusterized situations. In Appendix we give details
about this simulation scheme. First of all, we have verified the
stochastic approximation done in eq.~(\ref{velocidad}), using a simple
velocity field given by $u_x(\mathbf{r}_i(t),t)=U\cos(2\pi y_i(t)/L)$,
$u_y=0$, with $\mathbf{r}_i(t)=(x_i(t),y_i(t))$.  We calculate the
quantity given by the l.h.s of (\ref{velocidad}) minus the first three
terms in the r.h.s, cumulated for a little time period
$\overline{\Delta t}$ (greater than the simulation time step, but
shorter than the mean free time) and divided by $\tau_p^2
\sqrt{n}$. We have performed this for both cases: large and small $\tau_p$,
in order to check also the discussion below eq.~(\ref{velocidad}). 
One can appreciate in figure~\ref{fig:stochastic} that in the two cases
this
quantity is a Gaussian white noise.

Typical patterns for the distribution of particles obtained with DSMC 
at large time
can be seen in Fig.~\ref{fig:patterns}. The flow we have used here
is the cellular  flow derived from
the streamfunction~\cite{gollub} $\psi (x,y,t)=U \sin(2\pi/L(x+B_0 \cos (\omega t)))
\sin(2\pi/L y)$, with $B_0$ and $\omega$ 
the amplitude and frequency of the temporal perturbation,
respectively.
In this figure we just show a small part, around $2\%$ 
of the entire system, in
order to better appreciate the clustering areas. For the sake of clarity
we also show, in red lines,
the streamlines of the flow for $B_0=0$.  
 We have chosen two opposite situations. In figures~\ref{fig:patterns}A and 
~\ref{fig:patterns}B
we have studied a strong inertia case ($\tau_p>1$) with the panel
~\ref{fig:patterns}B corresponding to the same case as~\ref{fig:patterns}A
 but without collisions.
Note that in the 
absence of collisions $b>a$ and preferential concentration can be
recognized (panel~\ref{fig:patterns}B),
 while with (elastic) collisions we have that $c
\gg d \gg b \gg a$ so that diffusion dominates 
and the distribution of particles in~\ref{fig:patterns}A is homogeneous.
 Thus, in this case
we observe the dispersing effect of the collisions.
In figures~\ref{fig:patterns}C and~\ref{fig:patterns}D, instead, we have considered a
weak inertia case, $\tau_p <1$, where~\ref{fig:patterns}D
 is the case with
no collisions. Now, without collisions we have that  $a>b$ and
the flow homogenizes the distribution of grains (see~\ref{fig:patterns}D).
 With collisions,
instead, we have $d>a>b>c$ and collision induced clustering is
observed (\ref{fig:patterns}C). Therefore, this case shows the clustering effect of the
collisions. 

\begin{figure}[htb]
\begin{center}
\includegraphics[clip=true,width=.45\columnwidth, keepaspectratio]{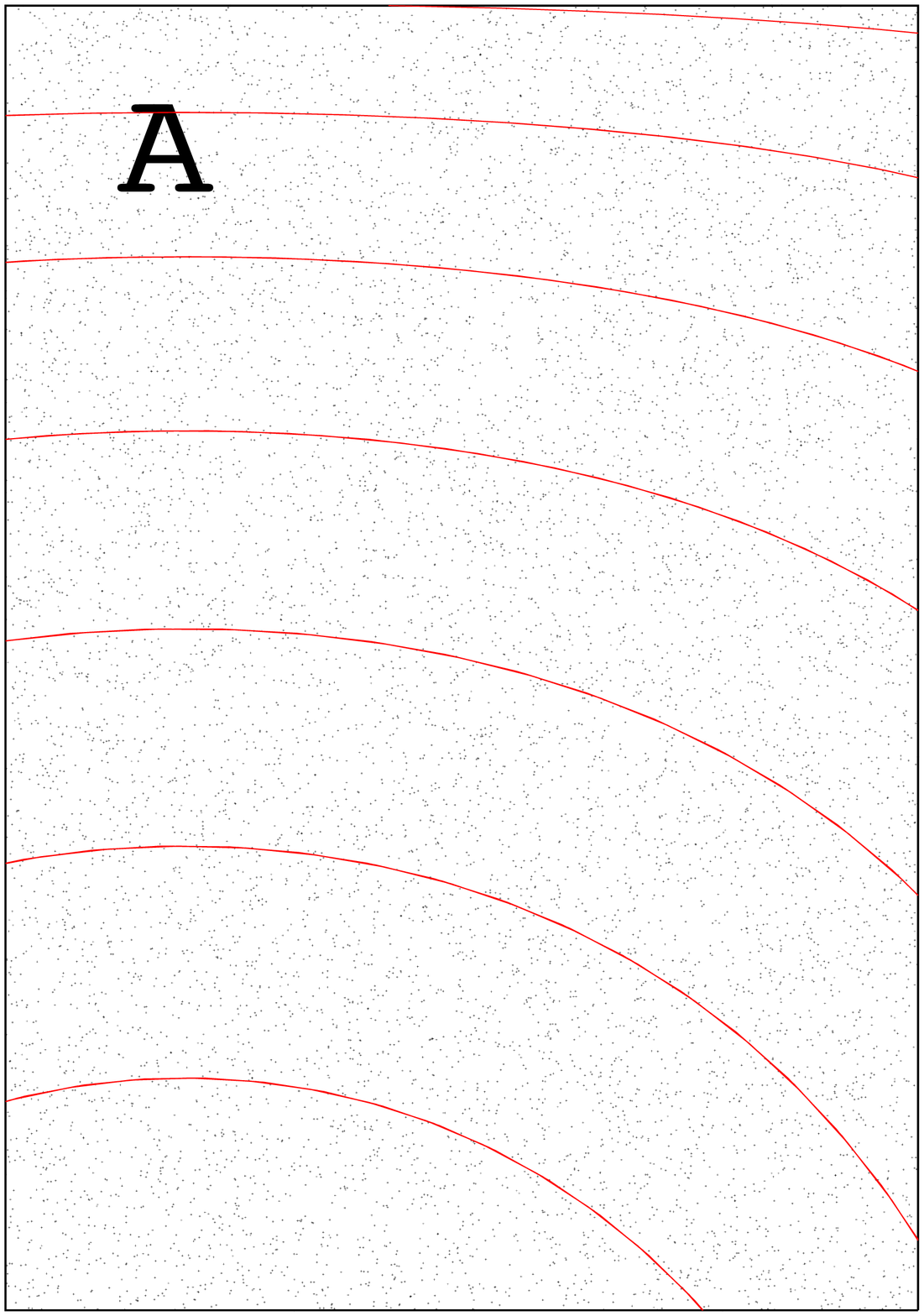}
\includegraphics[clip=true,width=.45\columnwidth, keepaspectratio]{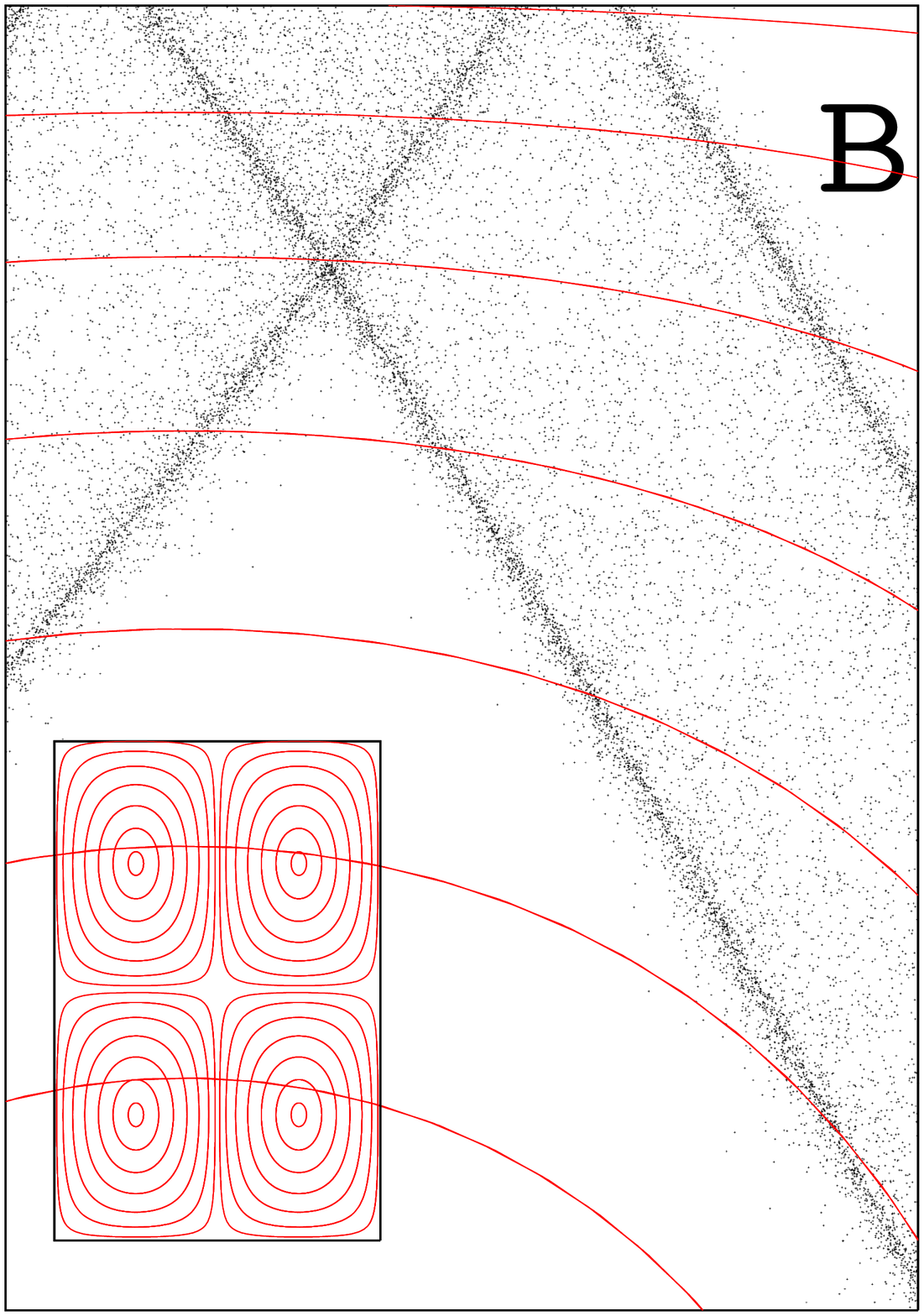}\\
\includegraphics[clip=true,width=.45\columnwidth, keepaspectratio]{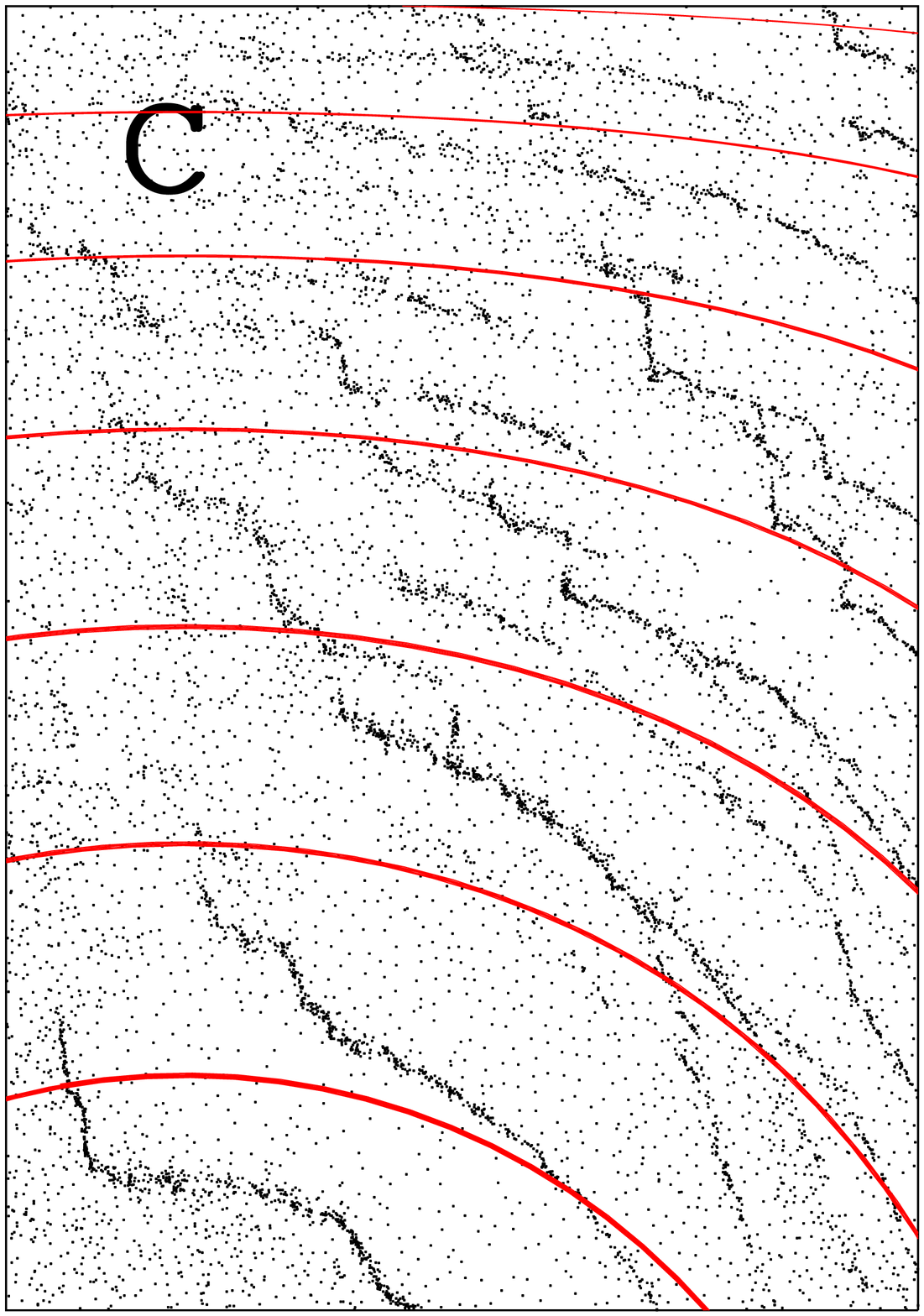}
\includegraphics[clip=true,width=.45\columnwidth, keepaspectratio]{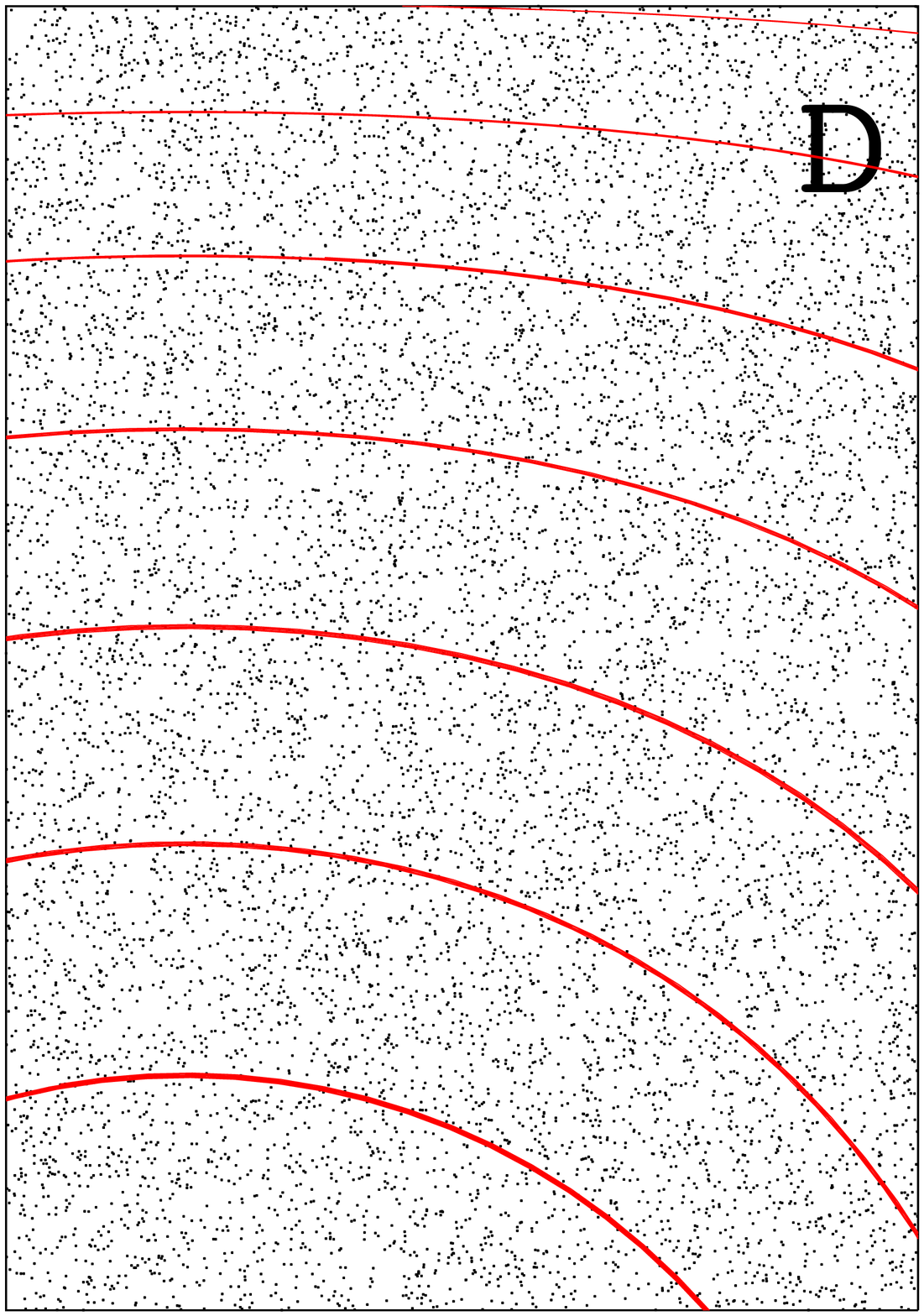}
\end{center}
\caption{ Instantaneous density patterns, obtained with DSMC,
for  the $2d$ cellular
flow;  $N=500000$ and
$L=\sqrt{N}$. B and D are without collisions. 
 Frames A, and B: $U=30$, $B_0=1$ $\omega=1$ ($\tau_f \sim
0.4$), $\hat{\tau_p}=10$, $\sigma=0.3$, $r=1$, $\tau_c \sim 0.02$ (only for A).
Figures
C and D: $U=110$, $B_0=1$, $\omega=10$ ($\tau_f \sim 0.3$),
$\tau_p=0.003$, $\sigma=0.5$, $r=1$, $\tau_c \sim 0.0002$ (for C). 
come from DSMC. on
Note that a small part ($\sim 2\%$)
of the whole system is portraid. Red lines represent streamlines of
the flux with $B_0=0$. The inset in B shows these
for the entire spatial domain.}
\label{fig:patterns}
\end{figure} 

The  numerical comparison of the DSMC with the continuum equation
is shown in Fig.~\ref{fig:clustering}. Here we measure the $P_M(n)$ function,
which gives the histogram (normalized to unity) of the number of boxes containing
$n$ particles after dividing the system in $M$ boxes. Note that as the clustering
is stronger the deviations of $P_M(n)$ from a Poissonian (homogenous distribution)
are more evident. On the right we plot the function for the DSMC patterns of
figs.~\ref{fig:patterns}A and \ref{fig:patterns}B (upper right panel) and 
figs.~\ref{fig:patterns}C and \ref{fig:patterns}D 
(lower right). On the left, we show the $P_M(n)$ for 
the distribution calculated with the continuum equation and the same value of the parameters.
Note that now a homogenous distribution of particles corresponds to a constant density and
that is  the reason why in this
 case the $P_M(n)$ resembles a Dirac Delta distribution centered
around this constant value.

The results shown in figs.~\ref{fig:patterns} and \ref{fig:clustering}
clearly indicate the relevant role of the collisions for inertial particles
immersed in a flow, verifying as well the important and
unexpected phenomenon of {\it reversed clustering} that appears on
the basis of the inspection of terms $c$ and $d$ of
equation~(\ref{final}). Briefly, in the absence of collisions
particles with inertia tend to aggregate, and the largest is the value
of $\tau_p$ the more compact is the aggregation of
particles~\cite{general}. This can be seen comparing
figs.~\ref{fig:patterns}B and \ref{fig:patterns}D.
  However, when collisions are taken
into account this last statement can be wrong, and just the contrary
occurs: inertia is decreasing but aggregation increases (again this is
what happens if we compare figs.~\ref{fig:patterns}A and \ref{fig:patterns}C).
  Note
that in these last two cases we have that $\tau_c < \tau_p$ so that collisions {\em
dominate} the dynamics.  Equation~(\ref{final}) perfectly
reflects this situation: the main difference between the two cases is
originated by the value of $\tau_p$ that influences the relative
importance of terms $c$ and $d$, as well as the relative importance
between $a$ and $b$ in the absence of collisions.

It is important to note that this effect may also appear in
segregation proccesses, i.e., in systems with particles of different
sizes or densities.  Obviously, these particles have different values
of $\tau_p$ and the same scenario just commented emerges.  Finally we
put in evidence that the inelasticity of collisions (typical of
granular materials) here  just plays the role of slightly enhance spatial
correlations, i.e. clustering. This is not taken into account by
equation~(\ref{final}) but can be observed in the DSMC simulations.

\begin{figure}[htb]
\begin{center}
\includegraphics[clip=true,width=9.0cm, keepaspectratio]{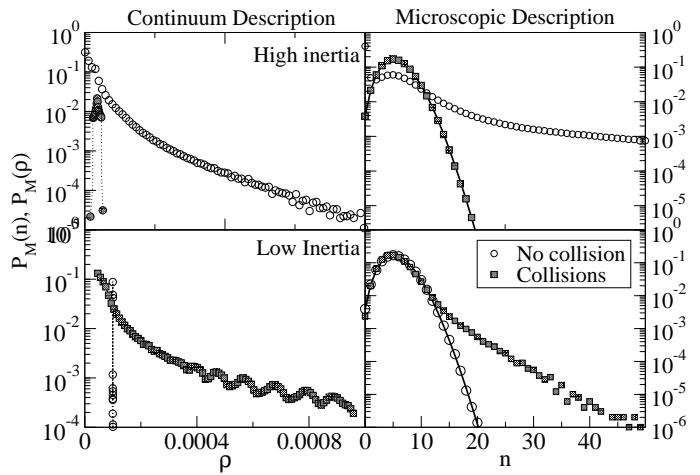}
\end{center}
\caption{$P_M(n)$ function for the distribution of particles obtained from DSMC (right
panels) and for the density of particles from the continuum equation~(\ref{final}) (left)
for the same parameter values. Upper right are the cases in~\ref{fig:patterns}A and 
\ref{fig:patterns}B, that is, with high inertia, and the lower right corresponds to the
\ref{fig:patterns}C and \ref{fig:patterns}D panels (low inertia). 
}
\label{fig:clustering}
\end{figure} 

\section{Summary}
\label{sec:summary}

The evolution of the density field of
 a large number of colliding finite-size 
particles driven by an external flow has
been derived. 
It is important to note that no premise has
been perfomed on the flow
so our results are equally valid for 
laminar, chaotic or turbulent flows. However, in the numerics
we have used a chaotic flow just for its interest and simplecity, and
also to compare with the results of our previous work~\cite{pre}.
In this equation one
can recognize two, in principle, 
competing roles of the collisions: clustering
and dispersion. The relative strengths of the 
important time scales involved in
the system  will finally determine if the particles
tend to cluster or, on the contrary,
are uniformly distributed in the space.
Collisions may even invert 
the typical scenario found
for inertial particles, showing the reversed
clustering phenomenon. We have also presented a numerical study
of the continuum equation and compared with DSMC simulations of
the system of discrete particles. The good agreement gives support
to our theory.

\section{Acknowledgments}

We acknowledge fruitful discussions with 
Umberto Marini Bettolo Marconi.
C.L. acknowledges support from MCyT of Spain
under project
REN2001-0802-C02-01/MAR (IMAGEN).

\section*{Appendix: the DSMC}

Direct Simulation Monte Carlo, also known as Bird Method~\cite{dsmc},
is a simulation scheme used to solve Boltzmann equations. With some
conditions and in well defined limits its results are proved to
converge to the solution of the Boltzmann equation for a gas of hard
spheres~\cite{matematici}. The algorithm consists in two main steps:
a) the free flow, and b) the collisions. Every time step of length $\Delta
t$ (lesser than any characteristic physical time, e.g. the collision
time) all the particles are moved {\it freely}, i.e. disregarding
possible collisions and then the collision procedure is applied: the
system is divided into cells of linear size lesser than any
characteristic physical length (e.g. the mean free path) and particles in
the same cell may collide {\it randomly}. For each cell $c$ a number
$m_c$ of couples of particles is chosen with probability proportional
to their scattering section, i.e. in this case proportional to their
relative velocities. The number $m_c$ is calculated as $m_c=\omega_c
\Delta t$, where $\omega_c$ is the average collision frequency
estimated in the cell $c$, by assuming a Gaussian distribution of
velocities with variance given by the actual variance of velocities of
the particles inside the cell. As it is known, the Boltzmann equation
is a correct description of a gas of hard particles only in the dilute
limit $N \to \infty$, $\sigma \to 0$ with finite $N \sigma^{d-1}$ ($d$
the space dimension). When the gas cannot be considered dilute,
correlations arise in the form of an enhancement of the collision
frequency and in complicated excluded volume effects. It is accepted
that at not too high packing fraction, the so-called Enskog correction
to the Boltzmann equation gives a sufficient description of these
effects. In the Boltzmann equation this correction appears as a simple
multiplicative term in front of the collision integral, which is
equivalent to an increase of the collision frequency. The Enskog
correction is usually taken to be the static correlation function
$g(r)$ evaluated at contact, i.e. $r=\sigma$, for which approximated
forms (dependent upon the local volume fraction $\phi=N_c\pi
(\sigma/2)^2/V_c$, with $V_c$ the area of a cell and $N_c$ the number
of particles in the cell) are available. We (in $d=2$) have used the
following form~\cite{correlation}:

\begin{equation}
g(\sigma)=\frac{1-7\phi/16}{(1-\phi)^2}
\end{equation}


\begin{thebibliography}{99}
\bibitem{dosdim} In two spatial dimensions the integral in the collision term can be 
performed 
and one obtains $ -\frac{\gamma \sigma^2 \tau_p
\pi}{4\tau_c}[\partial_x (\rho^2 (\partial_x u_y +
\partial_y u_x)+\partial_y(\rho^2(\partial_x u_x -\partial_y u_y))]$,
\end{thebibliography}
\end{document}